\documentclass[conference,a4paper]{IEEEtran}
\usepackage[left=1.62cm,right=1.62cm,top=1.9cm]{geometry}
\usepackage{balance}
\usepackage{float}

\usepackage{graphicx}
\usepackage{subfig}
\usepackage{tabularx}

\usepackage{standard}
\usepackage{etex}

\usepackage{rotate}
\usepackage{algorithmic}
\usepackage{algorithm}
\usepackage{float}
\usepackage{tikz}
\usepackage{refcount}

\usepackage{pgf,tikz}
\usepackage{pgfplots}
\usepackage{pgfplotstable}
\usepackage{bbm}
\usepackage{adjustbox}
\usetikzlibrary{calc}
\usepackage{relsize}
\usepackage{ifthen}

\usetikzlibrary{calc,fit,arrows,plotmarks,intersections,patterns,shapes,decorations,positioning, backgrounds, matrix}

\usepackage[numbers,sort&compress]{natbib}
\usepackage[ngerman, english]{babel}

\usetikzlibrary{arrows.meta}
\edef\wdArrowLength{2}
\tikzset{>={Latex[width=1.5mm,length=\wdArrowLength mm]}}
\usepackage{graphicx}

\usepackage{cancel}
\usepackage{mathtools}
\usepackage{shadethm}
\usepackage{empheq}
\usepackage{skull}
\usepgfplotslibrary{units}
\usepackage{calc}
\usepackage{nicefrac}
\usepackage{fancybox}
\usepackage{psfrag}
\usepackage{dsfont}

\title{Uncertainty Propagation and Minimization for Channel Estimation in UAV-mounted RIS Systems
\thanks{This work was supported in part by the German Research Foundation (DFG) in the course of the project SPP2433 under the project no. 541021107 (Measurement Technology on Flying Platforms) under grant SE 1697/22-1. This work was funded by the European Commission under the grant
no. 101079342 (Fostering Opportunities Towards Slovak Excellence in
Advanced Control for Smart Industries) and by the Deutsche Forschungsgemeinschaft
(DFG) under grant MO 1086/17-1 and grant INST 213/1028-1
FUGB.}} 
\author{\author{Kevin Weinberger$^{\dagger }$, David Müller$^{\ast }$, Martin M\"onnigmann$^{\ast }$, Aydin Sezgin$^{\dagger }$, \\
		$^{\ast }$Automatic Control and Systems Theory, Ruhr-Universit\"at Bochum, Germany,\\
        $^{\dagger }$Institute of Digital Communication Systems, Ruhr-Universit\"at Bochum, Germany,\\
		 \{Kevin.Weinberger, David.Mueller-r21, Martin.Moennigmann, Aydin.Sezgin\}@rub.de  }}
\date{\today}

\usepackage{graphicx}
\usepackage{placeins}

\usepackage{booktabs}
\usepackage{marvosym}
\usepackage{multirow}

\usepackage{latexsym, amsmath, amssymb, amsfonts, upgreek}

\usepackage[nolist]{acronym}

\usepackage{tikz}
\usetikzlibrary{calc,arrows,positioning,decorations,shadows,shapes,fadings,matrix}
\tikzset{>=latex'}
\tikzset{semithick}

\makeatletter
\providecommand{\IfElsePackageLoaded}[3]{\@ifpackageloaded{#1}{#2}{#3}}
\makeatother

\makeatletter

\def\tikz@delimiter#1#2#3#4#5#6#7#8{%
	\bgroup
		\pgfextra{\let\tikz@save@last@fig@name=\tikz@last@fig@name}%
		node[outer sep=0pt,inner sep=0pt,draw=none,fill=none,anchor=#1,at=(\tikz@last@fig@name.#2),#3]
		{%
			{\nullfont\pgf@process{\pgfpointdiff{\pgfpointanchor{\tikz@last@fig@name}{#4}}{\pgfpointanchor{\tikz@last@fig@name}{#5}}}}%
			\delimitershortfall\z@
			\resizebox*{!}{#8}{$\left#6\vcenter{\hrule height .5#8 depth .5#8 width0pt}\right#7$}%
		}
		\pgfextra{\global\let\tikz@last@fig@name=\tikz@save@last@fig@name}%
	\egroup%
}

\tikzset{hexagon/.code={
	\draw (0,2) -- (-4,0) -- (0,-2) -- (4,0) -- (0,2);
}}

\tikzset{phone/.code={
   \node [rectangle,rounded corners=1.5pt,draw,minimum height=0.6cm, minimum width=0.35cm] at (0,0){};
   \node [rectangle,rounded corners=1.5pt,draw,minimum height=0.5cm, minimum width=0.3cm] at (0,0){};
}}

\makeatletter
\def\cantox@vector#1#2#3#4#5#6#7#8{%
  \dimen@.5\p@
  \setbox\z@\vbox{\boxmaxdepth.5\p@
   \hbox{\kern-1.2\p@\kern#1\dimen@$#7{#8}\m@th$}}%
  \ifx\canto@fil\hidewidth  \wd\z@\z@ \else \kern-#6\unitlength \fi
  \ooalign{%
    \canto@fil$\m@th \CancelColor
    \vcenter{\hbox{\dimen@#6\unitlength \kern\dimen@
      \multiply\dimen@#4\divide\dimen@#3 \vrule\@depth\dimen@\@width\z@
      \vector(#3,-#4){#5}%
    }}_{\raise-#2\dimen@\copy\z@\kern-\scriptspace}$%
    \canto@fil \cr
    \hfil \box\@tempboxa \kern\wd\z@ \hfil \cr}}
\def\bcancelto#1#2{\let\canto@vector\cantox@vector\cancelto{#1}{#2}}
\makeatother

\newcommand{\ifthen}[2]{\ifthenelse{#1}{#2}{}}

\definecolor{myblue1}{rgb}{0,0,255}
\definecolor{myblue2}{rgb}{65,105,225}
\definecolor{myblue3}{rgb}{70,130,180}
\definecolor{myblue4}{rgb}{176,196,222}

\newcommand{\mytilde}{{\raise.17ex\hbox{$\scriptstyle\mathtt{\sim}$}}}
\newcommand{\naive}{}
\def\naive/{na\"{\i}ve}

\newcommand{\executeiffilenewer}[3]{%
\ifnum\pdfstrcmp{\pdffilemoddate{#1}}%
{\pdffilemoddate{#2}}>0%
{\immediate\write18{#3}}\fi%
}

\pgfkeys{/pgf/fpu}
\pgfmathparse{16383+1}

\pgfkeys{/pgf/fpu=false}

\IEEEoverridecommandlockouts

\usepackage{pdfpages}

\begin{document}
\bstctlcite{IEEEexample:BSTcontrol}

\maketitle

\begin{abstract}
Reconfigurable Intelligent Surfaces (RIS) are emerging as a key technology for sixth-generation (6G) wireless networks, leveraging adjustable reflecting elements to dynamically control electromagnetic wave propagation and optimize wireless connectivity. By positioning the RIS on an unmanned aerial vehicle (UAV), it can maintain line-of-sight and proximity to both the transmitter and receiver, critical factors that mitigate path loss and enhance signal strength. The lightweight, power-efficient nature of RIS makes UAV integration feasible, yet the setup faces significant disturbances from UAV motion, which can degrade RIS alignment and link performance.
In this study, we address these challenges using both experimental measurements and analytical methods. Using an extended Kalman filter (EKF), we estimate the UAV's orientation in real time during experimental flights to capture real disturbance effects. The resulting orientation uncertainty is then propagated to the RIS's channel estimates by applying the Guide to the Expression of Uncertainty in Measurement (GUM) framework as well as complex-valued propagation techniques to accurately assess and minimize the impact of UAV orientation uncertainties on RIS performance. This method enables us to systematically trace and quantify how orientation uncertainties affect channel gain and phase stability in real-time. Through numerical simulations, we find that the uncertainty of the RIS channel link is influenced by the RIS's configuration.  Furthermore, our results demonstrate that the uncertainty area is most accurately represented by an annular section, enabling a 58\% reduction in the uncertainty area while maintaining a 95\% coverage probability.
\end{abstract}
	\vspace{-0.1cm}
\section{Introduction}
Reconfigurable Intelligent Surfaces (RIS) are envisioned to enable the demanding requirements of sixth-generation (6G) wireless communication networks by facilitating precise control over the propagation environment \cite{Renzo2}. Through adjustable reflecting elements, RIS can modulate the phase, amplitude, and frequency of an impinging signal, making it possible to enhance network objectives like resilience and open up entirely new applications. To achieve optimal results, both the configuration and placement of the RIS are crucial, especially given the multiplicative path loss of the transmitter (Tx)-RIS-receiver (Rx) link \cite{bjornson2019intelligent}. In challenging terrains with obstacles and variable heights, maintaining an optimal line-of-sight (LoS) path becomes complex \cite{UAVTRaj}. Deploying RIS on unmanned aerial vehicles (UAVs) has emerged as an effective solution, leveraging the UAV's ability to adjust RIS positioning to ensure LoS and to adapt to environmental changes \cite{RIS_UAV_overview_new}. With its lightweight and energy-efficient design, RISs are an ideal alternative to heavy, power-intensive base stations in UAV applications \cite{UAV_RIS}.

While mounting a RIS on a UAV increases flexibility, it also introduces new uncertainties caused by external disturbances and (the UAV's) internal sources of uncertainty. As the UAV compensates for these disturbances by modifying its orientation and position, the RIS-reflected path also shifts, impacting signal transmission \cite{UAV_Kev}. This ongoing adjustment requires real-time RIS reconfiguration, as a fixed setup is insufficient even during stationary hover flight \cite{UAV_Kev}. To manage these dynamics, we leverage onboard sensors and an extended Kalman Filter (EKF) to estimate the UAV’s orientation and adapt the RIS configuration accordingly. However, as the EKF is subject to measurement errors and linearization, uncertainties persist in the orientation estimates, which we quantify to assess their influence on channel quality.

Additionally, we apply the Guide to the Expression of Uncertainty in Measurement (GUM) framework \cite{JCGMGUM} to rigorously propagate the measurement uncertainties within our model, allowing us to precisely study their impact on the effective channel link quality provided by the RIS. Using flight data from real-world experiments, we validate the RIS’s performance and show that the RIS configuration is actively influencing the uncertainty.

\section{System and Channel Model}\label{sec:sysmod}
\subsection{Experimental Setup}

\begin{figure}[]
	\centering
	\includegraphics[trim={2.2cm 1cm 0.25cm 1.5cm},clip,width=0.6\linewidth] {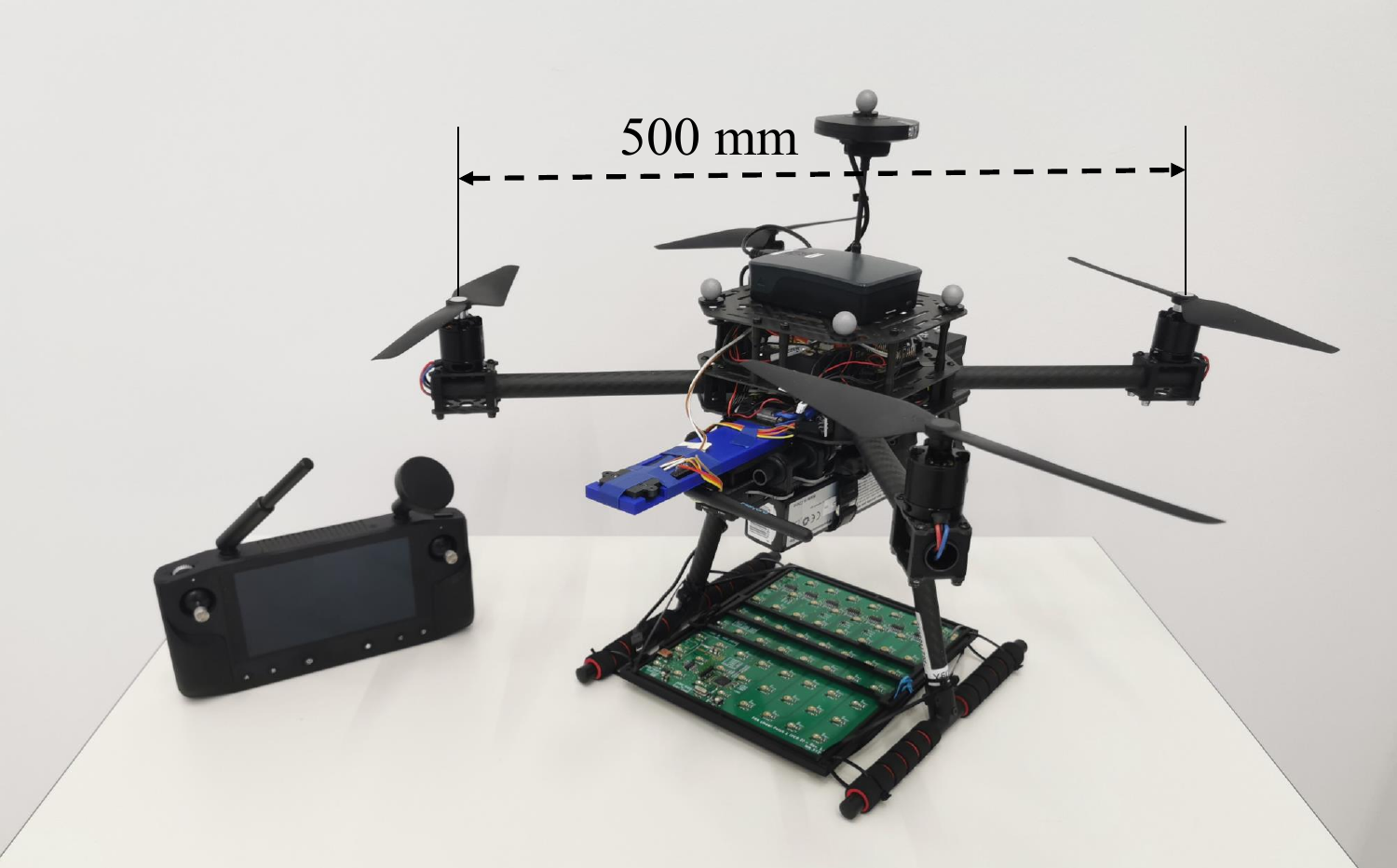} 
	\caption{\small Customized Holybro X500 equipped with RIS prototype and Herelink 1.1 controller.}
	\label{fig:UAV}\vspace{-0.2cm}
\end{figure}

The UAV carrying the RIS is a customized Holybro X500 Quadcopter (500 mm rotor-to-rotor) as shown in Fig \ref{fig:UAV}, able to carry up to 1 kg. The UAV is controlled by a Pixhawk Cube Orange flight controller running the open source software ArduPilot. The flight controller holds three different inertial measurement units (IMUs) (ICM20602, ICM20948, and ICM 206491), used for redundancy. ArduPilot is running an individual EKF for every IMU enabling lane-switching, a transition between multiple EKFs in case variances exceed a certain threshold. Remaining sensors are disabled for indoor use, with the altitude being measured by an offset-mounted VL53L1X laser sensor (200 mm offset from the UAV’s center). We adjusted the mass of the UAV in the model used by ArduPilot and tuned filter parameters for better filtering of vibrations. Reference orientation data is collected using a Vicon Vantage motion capture system (MCS) (10 V5 cameras, Tracker 3 software, $<$0.2 mm RMSE, 420 Hz). A RIS prototype \cite{OpenRIS} composed of $12\times10$ reflecting elements, each of which capable of individually applying $0^\circ$ and $180^\circ$ phase shifts, is mounted horizontally 30 cm below the UAV's center of gravity.  A Raspberry Pi, which is also attached to the UAV, serves as the RIS
controller. More information about the setup can be found in our previous work \cite{UAV_Kev}.

The experiments were conducted in a controlled lab environment with 10 hover flights (30 s each) at ~1 m altitude, using an altitude hold mode while position was controlled manually. The UAV was restarted and the IMUs calibrated before each flight to standardize initial conditions. The MCS was calibrated after warm-up and set to 100 Hz for consistent data logging with the flight controller. EKF data was logged at the same logging frequency as the MCS, i.e., 100 Hz, on an SD card. The UAV’s four-cell battery voltage ranged from 15.2 V to 16 V, which did not affect performance.
\vspace{-0.2cm}

\subsection{Data Prepocessing} Data from the motion capture and UAV were synchronized to eliminate shifts in the time series. The initial and final 5 seconds of each flight were removed (takeoff and landing), and a random 10-second interval (1000 data points) from the remaining 20 seconds was selected for evaluating orientation uncertainty.

\subsection{Channel Model}
The UAV carrying the RIS (e.g. Fig. \ref{fig:UAV}) is assumed to hover in the air, with the reflecting elements pointing towards the ground. There, we assume a single-antenna transmitter (Tx) transmitting signals towards the RIS prototype \cite{OpenRIS} totaling \( M=120 \) reflecting elements. The reflected signals from the RIS are then received by a single-antenna receiver (Rx). Due to the RIS's capability of inducing phase shifting to the reflected signals, the effective channel between Tx-RIS-Rx can be expressed as follows
\vspace{-0.25cm}
\begin{align}\label{eq:heff}
    h^\mathsf{eff} = \sqrt{G_{T}}\sqrt{G_{R}} \sum_{m=1}^{M} h_m \phi_m g_m,\\[-20pt]\nonumber
\end{align}
Here, the channel coefficient between the transmitter and the \( m \)-th RIS element is represented by \( h_m \), while \( g_m \) denotes the channel coefficient between the receiver and the \( m \)-th RIS element. The antenna gains at the transmitter and receiver are given by \( G_{T} \) and \( G_{R} \), respectively. The phase shift induced by the \( m \)-th RIS element on the reflected signal is expressed as \( \phi_m = e^{j\varphi_m} \), where \( \varphi_m \) represents the imposed phase shift.

With the above considerations, the cascaded Tx-RIS-Rx channel over the \( m \)-th RIS element is formulated as\vspace{-0.1cm}
\begin{align}
     h^\mathsf{casc}_m = h_m g_m,\\[-17pt]\nonumber
\end{align}
where \( h_m g_m \) represents the cascaded channel component.

In order to configure the phase shifts at the RIS optimally, knowledge of the cascaded Tx-RIS-Rx channels is required. To this end, we utilize the setup's current geometry. As shown in our previous works \citep{RIS_chanMod,ProtVal,DataSet,showmetheway}, optimizing the RIS prototype in this way yields RIS configurations in real-time, which result in performance improvements similar to experimentally obtained configurations. The cascaded channel components can be analytically determined for each reflect element \( m \) if the positions of Tx, RIS elements and Rx are known. We define \( d^h_m \) (or \( d^g_m \)) as the distance between the transmitter (or receiver) and the \( m \)-th element of the RIS. The cascaded Tx-RIS-Rx channel through the \( m \)-th RIS element is given by \cite{showmetheway}:\vspace{-0.1cm}
\begin{align}\label{h_casc}
     h^\mathsf{casc}_m = h_m g_m = \left[ \frac{c}{4\pi f d^h_m} e^{j\frac{2\pi}{\lambda}d^h_m} \right] \left[ \frac{c}{4\pi f d^g_m} e^{j\frac{2\pi}{\lambda}d^g_m} \right],
\end{align}
where \( c \) denotes the speed of light, \( \nu \) is the frequency, and \( \lambda \) represents the signal wavelength. Note that the antennas are in the RIS near-field, but the channel model remains valid as the individual reflect elements are not.

\section{Uncertainty of the UAV's orientation}\label{sec:exp}
We obtain the EKF estimation uncertainity for the UAV's orientation by comparing the EKF data to the values given by the MCS. The
difference of the reference roll, pitch and
yaw angles measured with the motion capture system to the
angles estimated with the EKF can be characterized by following the \textit{Guide to
the Expression of Uncertainty in Measurement (GUM)} \cite[Sec 4.2]{JCGMGUM}. More precisely, we can define $\Phi^\text{e}_k$, $\Theta^\text{e}_k$  and $\Psi^\text{e}_k$ as the absolute error of the roll, pitch and yaw angle, respectively, at time instance $k$.
Following the GUM, we can express the uncertainty in the orientation estimation as
the arithmetic mean \vspace{-0.2cm}
\begin{align}\label{q1}
    \Bar{q} = \frac{1}{n}\sum_{k=1}^{n} q_k,\\[-22pt] \nonumber
\end{align}
the experimental variance of observations \vspace{-0.1cm}
\begin{align}\label{q2}
    u^2(q_k) = \frac{1}{n-1}\sum_{j=1}^{n} (q_j-\Bar{q})^2,\\[-22pt] \nonumber
\end{align}
the experimental standard deviation $u(q_k)$,
the variance of the mean \vspace{-0.2cm}
\begin{align}\label{q3}
    u^2(\Bar{q}) = \frac{u^2(q_k)}{n},\\[-20pt] \nonumber
\end{align}
and the experimental standard deviation of the mean $u(\Bar{q})$ for the three quantities of interest $q_k= \Phi_k^\text{e}$, $\Theta_k^\text{e}$ and $\Psi_k^\text{e}$.

\section{Propagation of the Uncertainty}
The goal of this paper is to propagate the uncertainty values of the UAV's orientation angles estimated by the EKF $q\in\{\Phi,\Theta,\Psi\}$ for roll, pitch and yaw in order to obtain an uncertainty value for the effective Tx-RIS-Rx channel $h^\mathsf{eff}$. To this end, we first propagate the orientation uncertainty to the distance values $d^h_m$ and $d^g_m$ for the Tx-RIS and RIX-Rx channel, respectively. Next, the uncertainty of the distance values is propagated to the amplitude and phase of each Tx-RIS-Rx channel that is provided by the $m$-th reflecting element. Up to this point, the \textit{Law of Propagation of Uncertainty} (LPU) provided in the GUM can be utilized for propagation of the uncertainty values. However, the effective channel $h^\mathsf{eff}$ is a superposition of complex quantities, which have been imposed with a controllable phase shift. For this complex-valued case, the GUM does
not accommodate any guidelines on the correct form of propagation. As it turns out, the LPU for real-valued quantities presented in the GUM has been shown to be a univariate
form of a more general multivariate procedure \cite{Gum_general}. As a result, extensions of the LPU have been formulated for complex-valued
uncertainty propagation \cite{Hall_complex}, which are utilized in this work for proper assessment of the uncertainty for the effective channel. In the following, the first subsection will discuss the propagation in the real-valued regime, following the LPU in the GUM \cite[Sec.5]{JCGMGUM}. The second subsection will then explain how propagation moves from the real-valued to the complex-valued domain, and how it continues within the complex domain.

\subsection{Law of Propagation of Uncertainty}
An arbitrary measurement function\vspace{-0.15cm}
\begin{align}\label{eq:measFun}
y = f(x_1,x_2,...,x_N), \\[-20pt] \nonumber
\end{align}
describes a real-valued quantity of interest $y$, which depends on several influence quantities $x_1,x_2,...,x_N$. In this case, the quantity
$y$ is our measurand\textemdash the quantity we are interested in measuring or understanding. Based on \cite[Sec.5]{JCGMGUM} the combined standard uncertainty $u(y)$ is the positive square root of the combined variance $u^2(y)$, which is given by\vspace{-0.2cm}
\begin{align}\label{GumProp}
 u^2(y) = \sum_{i=1}^{N} \big(\underbrace{\frac{\partial f}{\partial x_i}}_{c_i}\big)^2 u^2(x_i), \\[-22pt]\nonumber
\end{align}
where each $u^2(x_i)$ is the standard uncertainty of the input quantities evaluated either statistically using datasets (Type A evaluation) or by being propagated (Type B evaluation) \cite[Sec.5]{JCGMGUM}. The partial derivative $c_i$ is also called sensitivity coefficient as it describes the impact on the output estimate in $y$, when $x_i$ is subjected to changes in the value of the input estimates. In the case of correlation between the input quantities, the appropriate expression for the combined variance expands to \vspace{-0.15cm}
\begin{align}\label{GumPropCorr}
 u^2(y) = \sum_{i=1}^{N} c_i^2 u^2(x_i) + 2 \sum_{i=1}^{N-1} \hspace{-0.1cm} \sum_{j=i+1}^{N} \frac{\partial f}{\partial x_i}\frac{\partial f}{\partial x_j} u(x_i,x_j),\\[-20pt] \nonumber
\end{align}

\subsection{Uncertainty Propagation to Amplitude and Phase}
We start by facilitating the propagation from the orientation uncertainty to the distance values for each reflecting element $m$ by denoting the position of the UAV's center of gravity as $\mathbf{p}_c$, the combined 3D rotational matrix as $\mathbf{R}(\Phi,\Theta,\Psi)$ and $\mathbf{r}_m$ as the vector describing the position from the reflective element $m$ relative to the UAV's center of gravity $\mathbf{p}_c$. Due to the assumption that the UAV only changes its orientation around its center of gravity the position of reflecting element $m$ can be formulated as
\begin{align}\label{eq:Pm}
  \mathbf{p}_m(\Phi,\Theta,\Psi) = \mathbf{p_c} + \mathbf{R}(\Phi,\Theta,\Psi)\mathbf{r}_m.
\end{align}
Since the rotation matrix comprises a sequence of rotations, it can be expressed as the product of the individual matrices for yaw, pitch and roll as \vspace{-0.1cm}
\begin{align}\label{eq:Rot}
\mathbf{R}(\Phi,\Theta,\Psi) = \mathbf{R}_z(\Psi)\mathbf{R}_y(\Theta)\mathbf{R}_x(\Phi).\\[-17pt] \nonumber
\end{align}
Given the fixed position of the transmitter $\mathbf{p}_\mathsf{Tx}$, the distance between Tx and RIS element $m$ can be formulated as
\begin{align}\label{dist}\vspace{-0.15cm}
  d^h_m = \norm{\mathbf{p}_m(\Phi,\Theta,\Psi) - \mathbf{p}_\mathsf{Tx}},
\end{align}
where (\ref{dist}) is now in the form of (\ref{eq:measFun}).

Using the chain rule, we calculate the sensitivity coefficients $c_q$ for $q\in\{\Phi,\Theta,\Psi\}$, i.e., for roll pitch and yaw, as \vspace{-0.15cm}
\begin{align}\label{sensD}
\hspace{-0.15cm} c_q^{d_m^h} \hspace{-0.075cm}=\hspace{-0.075cm}  \frac{\partial d_m^h}{\partial q}  \hspace{-0.05cm}\overset{(\ref{dist})}{=}\hspace{-0.05cm} \frac{ \mathbf{p}_m(\Phi,\Theta,\Psi) - \mathbf{p}_\mathsf{Tx}} {\norm{\mathbf{p}_m(\Phi,\Theta,\Psi) - \mathbf{p}_\mathsf{Tx}}}\frac{\partial}{\partial q}\mathbf{p}_m(\Phi,\Theta,\Psi),
\end{align}
where $q$ is one of the Euler angles. As the expression of  $\mathbf{p}_m$ in (\ref{eq:Pm}) only contains the orientation angles within the rotation matrix $\mathbf{R}$, the derivatives with respect to each angle can be calculated easily using the decomposition in (\ref{eq:Rot}). The calculations for the distance between RIS-Rx $d_m^g$ and their sensitivity coefficients follow the same procedure outlined above and are therefore omitted in this work.

With the above definitions, the standard combined uncertainty for the distances can be expressed as
\begin{align}\label{d_uncert}
  u^2(d_m^h) =& \sum_{q\in\{\Phi,\Theta,\Psi\}} (c_q^{d_m^h})^2 u^2(q), \\
  u^2(d_m^g) =& \sum_{q\in\{\Phi,\Theta,\Psi\}} (c_q^{d_m^g})^2 u^2(q).
\end{align}

In the context of GUM, accounting for the correlation between quantities that share common influencing variables is essential, even if those variables are themselves uncorrelated. Here, this indirect correlation between $d_m^h$ and $d_m^g$ arises from the shared dependency on the Euler angles $q$ and must be included to accurately quantify the combined uncertainty in measurements that are not independent by virtue of their mutual inputs \cite[Sec.5.2]{JCGMGUM}. Ignoring these correlations can lead to incorrect uncertainty estimates, while properly accounting for them improves the precision of any subsequent uncertainty propagation.
To this end, we capture the indirect correlation of $d_m^h$ and $d_m^g$ through their mutual dependence on the angles utilizing the covariance propagation formula \vspace{-0.1cm}
\begin{align}\label{cov_prop_form}
  u(y_1,y_2) =& \sum_{i=1}^{N}\sum_{j=1}^{N} \frac{\partial f_1}{\partial x_i}  \frac{\partial f_2}{\partial x_j} u(x_i,x_j),
\end{align}
resulting in \vspace{-0.1cm}
\begin{align}\label{u_dh_dg}
u(d_m^h,d_m^g) = \sum_{q\in\{\Phi,\Theta,\Psi\}} c_q^{d_m^h} c_q^{d_m^g} u^2(q),
\end{align}
since the Euler angles $q$ are assumed to be independent.

Further, we continue to propagate the uncertainty to the amplitude and phase values of the cascaded Tx-RIS-Rx channel over reflecting element $m$. To this end, we reformulate (\ref{h_casc}) as
\begin{align}\label{h_casc_prop}
  h_m^\mathsf{casc} = \frac{c^2}{(4 \pi \nu)^2  d_m^h d_m^g} e^{j \frac{2\pi}{\lambda} (d_m^h+d_m^g)}
\end{align}
such that we can easily obtain the formulations for the amplitude $A_m$ and phase $P_m$ values
\begin{align}
  A_m =& \frac{c^2}{(4 \pi \nu)^2  d_m^h d_m^g}, \label{Am}\\
  P_m =& \frac{2\pi}{\lambda}(d_m^h+d_m^g) \label{Pm}.
\end{align}
To facilitate the propagation to the amplitude and phase values, we determine the sensitivity coefficients as
\begin{align}
  c_{d_m^h}^{A_m} &= -\frac{c^2}{(4 \pi \nu d_m^h)^2   d_m^g},
  &&c_{d_m^g}^{A_m} = -\frac{c^2}{(4 \pi \nu d_m^g)^2   d_m^h},\\
  c_{d_m^h}^{P_m} &=  \frac{2\pi}{\lambda},
  &&c_{d_m^g}^{P_m} =  \frac{2\pi}{\lambda}.
\end{align}
To account for the indirect correlation between $d_m^h$ and $d_m^g$, we utilize (\ref{GumPropCorr}) to correctly propagate the standard combined uncertainty for $A_m$ and $P_m$ as
\begin{align}
  u^2(A_m) &=  (c_{d_m^h}^{A_m})^2 u^2(d_m^h) + (c_{d_m^g}^{A_m})^2
  u^2(d_m^g) \nonumber  \\
  &\qquad\qquad\qquad\qquad+ 2c_{d_m^h}^{A_m}c_{d_m^g}^{A_m} u(d_m^h,d_m^g)  \\
  u^2(P_m) &=  (c_{d_m^h}^{P_m})^2 u^2(d_m^h) + (c_{d_m^g}^{P_m})^2 u^2(d_m^g) \nonumber  \\
  &\qquad\qquad\qquad\qquad+ 2c_{d_m^h}^{P_m}c_{d_m^g}^{P_m} u(d_m^h,d_m^g).
\end{align}
It follows that this indirect correlation is also present between $A_m$ and $P_m$, which can be again be determined with (\ref{cov_prop_form}):
\begin{align}
  u(A_m,P_m) =& \, c_{d_m^h}^{A_m} c_{d_m^h}^{P_m} u^2(d_m^h) + c_{d_m^g}^{A_m} c_{d_m^g}^{P_m} u^2(d_m^g) \nonumber \\
  &+(c_{d_m^h}^{A_m} c_{d_m^g}^{P_m} + c_{d_m^g}^{A_m} c_{d_m^h}^{P_m} ) u(d_m^h,d_m^g).
\end{align}

\def\re{\mathcal{R}}
\def\im{\mathcal{I}}
\subsection{Uncertainty Propagation to Effective RIS Channel}
In this subsection, the real-valued amplitude and phase values for the uncertainty are propagated into the complex domain. This becomes necessary because the effective RIS channel $h^\mathsf{eff}$ is a superposition of complex values. In addition, we need to account for the RIS's ability to reconfigure its phase shift that is imposed on the reflection at each reflecting element $m$ individually.

To accomplish that, we briefly review the method of propagating covariance in \cite[Sec.2]{Hall_complex}. As such, we expand the measurement function (\ref{eq:measFun}) to also be applicable for \textit{complex-valued} inputs and outputs as
\begin{align}\label{y_comp}
  \mathbf{y} =& \mathbf{f}(\mathbf{X}) = \mathbf{f}(\mathbf{x}_1,\mathbf{x}_2,\dots, \mathbf{x}_N) = f_\re(\mathbf{X}) + j f_\im(\mathbf{X}),
\end{align}
where $\mathbf{y}_n=[y^{\re}, y^{\im}]^\mathsf{T}$ and  $\mathbf{x}_n=[x_n^{\re}, x_n^{\im}]^\mathsf{T}$ are vectors stacking the real and imaginary part of the output and input variables, respectively. Similarly, the function $\mathbf{f}$ comprises two scalar functions $f_{\re}$ and $f_{\im}$ evaluating the real and imaginary part, respectively.
The uncertainty in the values assigned to input quantities can then be represented by a \( 2N \times 2N \) covariance matrix,
\begin{align}
&\mathbf{U}(\mathbf{X}) =\\
&\begin{bmatrix}
u^2(x_{1}^\re) \!& \!u(x_{1}^\re,\! x_{1}^\im) & \cdots & u(x_1^\re,\! x_M^\re) & u(x_1^\re,\! x_M^\im) \\
u(x_1^\im,\! x_1^\re) & u^2(x_1^\im) & \cdots & u(x_1^\im,\! x_M^\re) & u(x_1^\im,\! x_M^\im) \\
\vdots & \vdots & \ddots & \vdots & \vdots \\
u(x_M^\re,\! x_1^\re) & u(x_M^\re,\! x_1^\im) & \cdots & u^2(x_M^\re) & u(x_M^\re,\! x_M^\im) \\
u(x_M^\im,\! x_1^\re) & u(x_M^\im,\! x_1^\im) & \cdots & u(x_M^\im,\! x_M^\re) & u^2(x_M^\im) \nonumber \\
\end{bmatrix},
\end{align}
where the diagonal terms, \( u^2(x_{m}) \), represent the standard variance of the respective inputs. Since $\mathbf{y}$ is decomposed in the real and imaginary parts, the uncertainty is expressed in a $2\times2$ covariance matrix, obtained using a bivariate form of the Gaussian error
propagation law. The law prescribes how uncertainty in
values assigned to the inputs propagates to an estimate of the
uncertainty in a measurement result. The covariance is written as
\begin{align}\label{ComplexCovPorp}
\mathbf{U}(\mathbf{y}) = \mathbf{J}(\mathbf{y})\mathbf{U}(\mathbf{X})\mathbf{J}(\mathbf{y})^\mathsf{T},
\end{align}
where $\mathbf{J}(\mathbf{y})$ is a $2\times2M$ Jacobian matrix with the following block structure
\begin{align}\label{Jacobian}
  \mathbf{J}(\mathbf{y}) &= [ \mathbf{J}_1(\mathbf{y}),\mathbf{J}_2(\mathbf{y}),\dots, \mathbf{J}_M(\mathbf{y})],
\end{align}
where\vspace{-0.2cm}
\begin{align}
 \displaystyle   \mathbf{J}_m =
    \begin{bmatrix}
    \dfrac{\partial f_\re}{\partial x_m^\re} \!& \!\dfrac{\partial f_\re}{\partial x_m^\im}\\[10pt]
    \dfrac{\partial f_\im}{\partial x_m^\re} \!& \!\dfrac{\partial f_\im}{\partial x_m^\im}\\
    \end{bmatrix},
\end{align}
in which each block represents the bivariate sensitivity coefficients of the problem.

Utilizing the definitions above, we can now propagate the real-valued uncertainty from the amplitude and phase into the complex domain, starting with the evaluation of $h_m^\mathsf{casc}$. To this end, we rewrite $h_m^\mathsf{casc}$ into the form of (\ref{y_comp}) as
\begin{align}
  h_m^\mathsf{casc} &= A_m e^{jP_m} = \underbrace{A_m \cos(P_m)}_{f_\re} +j \underbrace{A_m \sin(P_m)}_{f_\im}.
\end{align}
Because the input values are real-valued, the covariance matrix reduces to the form of
\begin{align}\label{Cov_AmPm}
&\mathbf{U}([A_m,P_m]) =
&\begin{bmatrix}
u^2(A_m) \!& \!0 & u(A_m,\! P_m) & 0 \\
0 & 0 & 0 & 0  \\
u(A_m,\! P_m) & 0 &  u^2(P_m) & 0 \\
0 & 0 &  0 & 0 \nonumber \\
\end{bmatrix}
\end{align}
and by denoting $\mathbf{h}_m^\mathsf{casc} = [\Re(h_m^\mathsf{casc}),\Im(h_m^\mathsf{casc})]^\mathsf{T}$, the Jacobian simplifies to
\begin{align}
 \displaystyle   \mathbf{J}(\mathbf{h}_m^\mathsf{casc}) \! = \![ \mathbf{J}_{A_m} \mathbf{J}_{P_m}]\! = \!\begin{bmatrix}
    \cos(P_m) \!& \!0 \!& -A_m\sin(P_m) \!& \!0 \\
    \sin(P_m) \!& \!0 \!& A_m\cos(P_m) \!& \!0\\
    \end{bmatrix}.
\end{align}
Utilizing (\ref{ComplexCovPorp}), the uncertainty of the real and imaginary values for $h_m^\mathsf{casc}$ can be determined as
\begin{align}\label{U_hm}
  \mathbf{U}(\mathbf{h}_m^\mathsf{casc}) &= \mathbf{J}(\mathbf{h}_m^\mathsf{casc}) \mathbf{U}([A_m,P_m]) \mathbf{J}(\mathbf{h}_m^\mathsf{casc})^\mathsf{T}\\
  &= \begin{bmatrix} u^{h_m^\textsf{casc}}_{11} & u^{h_m^\textsf{casc}}_{12} \\ u^{h_m^\textsf{casc}}_{21} & u^{h_m^\textsf{casc}}_{22} \end{bmatrix}, \\[-20pt] \nonumber
\end{align}
where
\begin{align}
    & u^{h_m^\textsf{casc}}_{11} = \cos^2(P_m) \, u^2(A_m) + A_m^2 \sin^2(P_m) \, u^2(P_m) \\&\qquad- 2 A_m \cos(P_m) \sin(P_m) \, u(A_m, P_m), \nonumber \\
    &u^{h_m^\textsf{casc}}_{22} = \sin^2(P_m) \, u^2(A_m) + A_m^2 \cos^2(P_m) \, u^2(P_m) \\&\qquad + 2 A_m \cos(P_m) \sin(P_m) \, u(A_m, P_m),\nonumber\\
    &u^{h_m^\textsf{casc}}_{12 \text{\textbackslash} 21} =(A_m (\cos^2(P_m) -\sin^2(P_m))) \, u(A_m, P_m) +\\& \sin(P_m) \cos(P_m) \, u^2(A_m) - A_m^2 \sin(P_m) \cos(P_m) \, u^2(P_m)
    \nonumber.
\end{align}

In the next and step in the chain of propagations, the uncertainty is propagated to the effective RIS channel over the $m$-th RIS element $h^\mathsf{eff}_m = h_m^\mathsf{casc} \phi_m$. For the derivation of the propagation we write the $m$-th effective channel over the RIS in the form of (\ref{y_comp}) as
\begin{align}\label{hEff}
  h^\mathsf{eff}_m = & \, \overbrace{\Big(\Re(h_m^\mathsf{casc}) \Re(\phi_m) - \Im(h_m^\mathsf{casc}) \Im(\phi_m) \Big)}^{f^\re}+\nonumber \\ &j\underbrace{(\Re\Big(h_m^\mathsf{casc}) \Im(\phi_m)+ \Im(h_m^\mathsf{casc}) \Re(\phi_m)\Big)}_{f^\im}.
\end{align}
The Jacobian, therefore, takes the form
\begin{align}
 \displaystyle   \mathbf{J}(\mathbf{h}_m^\mathsf{eff}) = \begin{bmatrix}
    \cos(\varphi_m) \!& -\sin(\varphi_m) \\
    \sin(\varphi_m) \!&  \cos(\varphi_m) \\
    \end{bmatrix},
\end{align}
which corresponds to a 2D rotation matrix, encapsulating the phase-shifting behavior of the RIS in the complex domain. The combined covariance of $\mathbf{h}_m^\mathsf{eff} = [\Re(h_m^\mathsf{eff}),\Im(h_m^\mathsf{eff})]^\mathsf{T}$ in the complex domain is then expressed as
\begin{align}\label{U_hm_eff}
  \mathbf{U}(\mathbf{h}_m^\mathsf{eff}) &= \mathbf{J}(\mathbf{h}_m^\mathsf{eff}) \mathbf{U}(\mathbf{h}_m^\mathsf{casc}) \mathbf{J}(\mathbf{h}_m^\mathsf{eff})^\mathsf{T}\\
  &= \begin{bmatrix} u^{h_m^\textsf{eff}}_{11} & u^{h_m^\textsf{eff}}_{12} \\ u^{h_m^\textsf{eff}}_{21} & u^{h_m^\textsf{eff}}_{22} \end{bmatrix}, \\[-20pt] \nonumber
\end{align}
where \vspace{-0.1cm}
\begin{align}
          u_{11}^{{h}_m^\mathsf{eff}} =& \cos^2(\varphi_m)u^{h_m^\textsf{casc}}_{11} + \sin^2(\varphi_m)u^{h_m^\textsf{casc}}_{22} \\ \nonumber & \qquad\qquad\qquad-2(\cos(\varphi_m)\sin(\varphi_m) u^{h_m^\textsf{casc}}_{21}),\\
           u_{22}^{{h}_m^\mathsf{eff}} =& \sin^2(\varphi_m)u^{h_m^\textsf{casc}}_{11} + \cos^2(\varphi_m)u^{h_m^\textsf{casc}}_{22} \\ \nonumber & \qquad\qquad\qquad+2(\cos(\varphi_m)\sin(\varphi_m) u^{h_m^\textsf{casc}}_{21}),\\
          u_{12 \text{\textbackslash} 21}^{{h}_m^\mathsf{eff}} =& \cos(\varphi_m)\sin(\varphi
          _m) u^{h_m^\textsf{casc}}_{11}  \\ \nonumber - \cos&(\varphi_m)\sin(\varphi) u^{h_m^\textsf{casc}}_{22} +(\cos^2(\varphi_m)-\sin^2(\varphi_m)) u^{h_m^\textsf{casc}}_{21}.
\end{align}

The final step involves propagating the combined uncertainty across the superposition of the $M$ effective channels ${h}_m^\mathsf{eff}$. Ignoring the antenna gains during the propagation process, we can write \vspace{-0.2cm}
\begin{align}
    h^\mathsf{eff} = \sum_{m=1}^{M} h_m^\mathsf{eff}. \\[-18pt] \nonumber
\end{align}
Since the input values $h_m^\mathsf{eff}$ can be considered independent, the Jacobian results in an identity matrix. For this case, the uncertainty can be propagated by summation of the $M$ matrices of the input values $\mathbf{U}(\mathbf{h}_m^\mathsf{eff})$ \cite{Hall_complex}. Thus, the combined uncertainty matrix for the effective channel is \vspace{-0.1cm}
\begin{align}\label{U_hEff}
   \mathbf{U}(\mathbf{h}^\mathsf{eff}) \! &= \!\begin{bmatrix} u_{11}^{h^\mathsf{eff}} \!\!\!&\!\!\! u_{12}^{h^\mathsf{eff}} \\ u_{21}^{h^\mathsf{eff}} \!\!\!&\!\!\! u_{22}^{h^\mathsf{eff}} \\ \end{bmatrix} \!=\!  \begin{bmatrix} \sum_{m=1}^{M} u^{h_m^\textsf{eff}}_{11}\! &\! \sum_{m=1}^{M} u^{h_m^\textsf{eff}}_{12} \\ \sum_{m=1}^{M}u^{h_m^\textsf{eff}}_{21} \!&\! \sum_{m=1}^{M} u^{h_m^\textsf{eff}}_{22} \end{bmatrix}.
\end{align}

\section{Simulation Results}
\begin{table}[bp] \small
    \centering
    \caption{Uncertainties in orientation estimation}
    \begin{tabular}{c c c c}
         \toprule
         GUM parameters & $\Bar{q}$ & $u(q_k)$ & $u(\Bar{q})$ \\
         \midrule
         $\Phi_e$ & 0.23{$^\circ$} & 0.49{$^\circ$} & 0.005{$^\circ$} \\
         $\Theta_e$ & 0.22{$^\circ$} & 0.48{$^\circ$} & 0.005{$^\circ$} \\
         $\Psi_e$ & -0.06{$^\circ$} & 0.18{$^\circ$} & 0.002{$^\circ$} \\
         \bottomrule
    \end{tabular}
    \label{tab:GUM_parameters}
\end{table}
\begin{table}[bp] \small
    \centering
    \caption{Success-rate within expanded uncertainty area}
    \begin{tabular}{c c c c}
         \toprule
          RIS configurations & $\varphi_{0}$ & $\varphi_{rand}$ & $\varphi_{opt}$ \\
         \midrule
         Elliptical Representation & 100\% & 100\% & 47.49\% \\
        Annular Representation & 100\% & 100\% & 99.79\% \\
         \bottomrule
    \end{tabular}
    \label{tab:Precision_parameters}
\end{table}
We assume a system that consists of a transmitter located at (0,0,0.1) m, a receiver positioned at (2,0,0.1) m, and the RIS placed at (0,0,0.7) m. The UAV's center of gravity, which is tracked by both the EKF system and MCS, is assumed to be located at (1,1,1) m. Note that due to the offset of the RIS's center point by 0.3 m from the UAV’s center of gravity, the reflecting elements follow a 3D parabolic trajectory, as they rotate around the UAV’s center of gravity. We further assume antenna gains of $G_T=G_R=1$. In order to determine the uncertainty values for the orientation angles, we utilize our measured dataset of the UAV's hover flights. Utilizing the definitions in (\ref{q1})-(\ref{q3}) to calculate the uncertainties in orientation estimation, the values listed in Table \ref{tab:GUM_parameters} can be calculated. These values are used as starting point for the uncertainty propagation. During the propagation process, we evaluate three RIS configurations:
1) $\varphi_{0}$ denoting the off state, $\varphi_m = 0, \forall m\in M$, where the RIS reduces to a metal reflector; 2) $\varphi_{rand}$ representing random RIS configurations at each data point; and 3) $\varphi_{opt}$, representing an optimized RIS, where $\varphi_m = -\arg(h_m^\mathsf{casc})$. 

%
\begin{figure}
  \centering
  \includegraphics[width=0.9\linewidth]{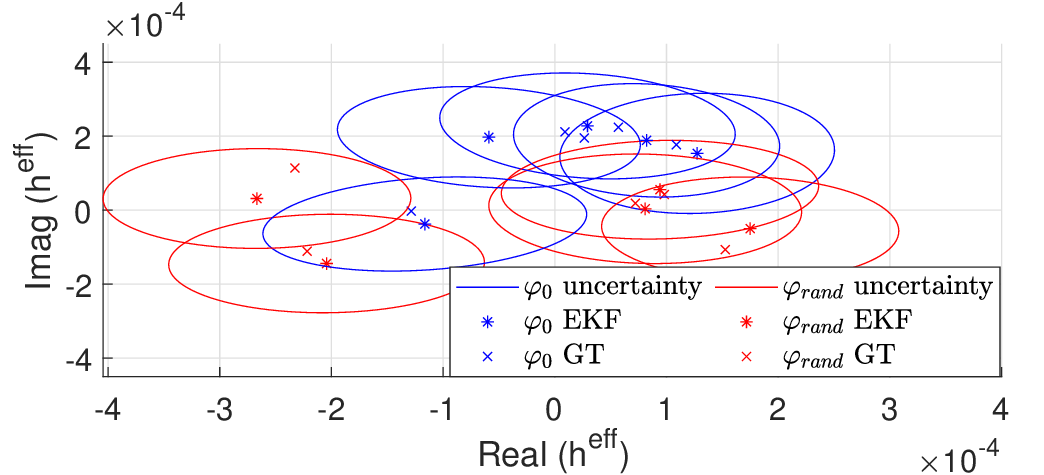}
  \caption{\small EKF estimated channels and their Ground Truth (GT) values alongside the resulting uncertainty ellipse for $\mathbf{U}(\mathbf{h}^\mathsf{eff}$) for the RIS states $\varphi_0$ (metal plate) and $\varphi_{rand}$ (random), respectively. \\[-21pt] }\label{fig:uncertRand}
\end{figure}
\begin{figure}
  \centering
  \includegraphics[trim={2cm 10.75cm 1.8cm 10.75cm},clip,width=0.9\linewidth] {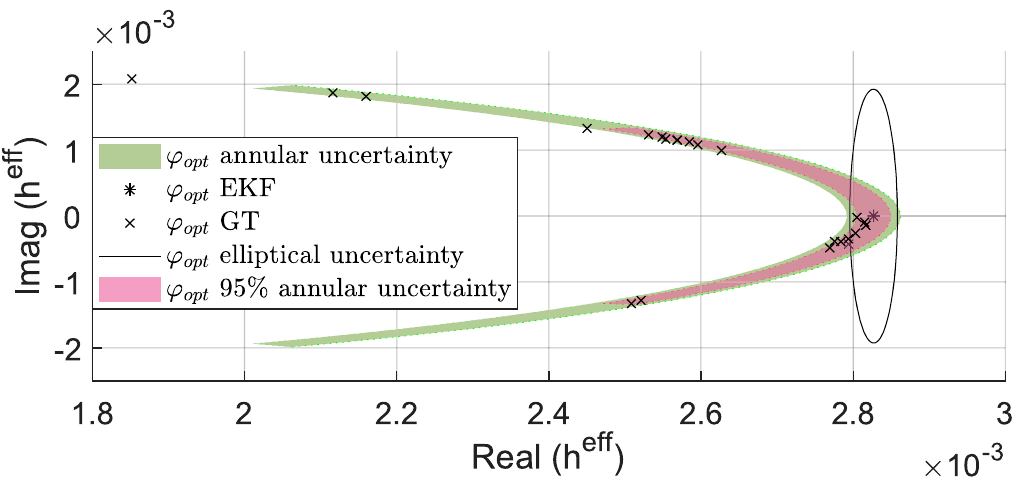}
  \caption{\small EKF estimated effective channel and its uncertainty ellipse as well as the associated annular section for the optimized RIS states $\varphi_{opt}$. Representative GT values and the annular section with 95\% coverage probability are also included.\\[-25pt]}\label{fig:uncertOpt}
\end{figure}


To analyze the results, we use the expanded uncertainty, which, as defined by the GUM, represents intervals within which a specified percentage of data points are expected to fall. However, applying the concept of expanded uncertainty to covariance matrices, such as $\mathbf{U}(\mathbf{h}^\mathsf{eff})$, is not straightforward. We solve this problem, by interpreting this uncertainty as elliptical regions in the complex domain, offering an accurate and practical representation \cite{Hall_2016}. As per GUM, we can construct a region with a level of confidence of approximately 95\%, using the coverage factor $k_{\mathsf{GUM},{0.95}} =2$, which is also often adapted in UAV and networking applications.

Fig. \ref{fig:uncertRand} displays the EKF estimated channels, their Ground Truth (GT) as well as their uncertainty ellipse, calculated from  $\mathbf{U}(\mathbf{h}^\mathsf{eff})$, using the RIS configurations $\varphi_0$ (metal plate) and $\varphi_{rand}$ (random), respectively. To improve visual clarity, only a subset of representative points from the dataset is shown. The figure shows that the uncertainty ellipses have different eccentricities and rotations. This directly results from the movement of the UAV, which alters the relative positioning of each RIS elements, therefore changing the channels, and thus the region of uncertainty, in each data point. To evaluate the precision of the propagated uncertainty, Table \ref{tab:Precision_parameters} presents the success-rate for these two RIS cases,  where a 'success' is defined as an ellipse encompassing a ground truth (GT) data point. With a 100\% success-rate this elliptic representation of $\mathbf{U}(\mathbf{h}^\mathsf{eff})$ seems feasible. However, the table also reveals a significantly reduced sucess-rate of 47.49\% when optimized RIS configurations are utilized.
To investigate this behaviour, Fig. \ref{fig:uncertOpt} shows the EKF channel estimation (for one representative data point) as well as its uncertainty ellipse. In addition more GT points from the dataset are showcased. Since the optimized RIS configuration shifts all individual channels' phases $h_m^\mathsf{casc}$ to the positive real axis, the EKF-estimated effective channels to have no imaginary part. However, the GT values of the channels $h_m^\mathsf{eff}$ lie within an elliptic area around the EKF values, making the EKF phase shifts suboptimal for the GT channels. This leads to phase misalignments, and as shown in the figure, the superposition of these misalignments introduces imaginary parts in the GT values, which the ellipse cannot account for.

With this insight, we propose to transform the uncertainty values in $\mathbf{U}(\mathbf{h}^\mathsf{eff})$ to annular sections as described in \cite{Hall_complex}, using the proposed coverage factor $k_{\mathsf{Ann},0.95} = 2.24$. Fig. \ref{fig:uncertOpt} demonstrates that the annular representation vastly improves the coverage area shape, capturing the uncertainty more accurately. As shown in Table \ref{tab:Precision_parameters}, this representation boosts the success-rate to 99.79\% for the optimized RIS case, while maintaining 100\% for the off and random RIS states. Additionally, by reducing the coverage factor to $k_{\mathsf{RIS},0.95} = 1.445$ to achieve a 95\% coverage probability, the uncertainty area is reduced by 58\% as shown in Fig. \ref{fig:uncertOpt}. Narrowing the region where the channel values can vary leads to a higher confidence in predicting where the GT values might lie.

\section{Conclusion}
This paper presents a rigorous uncertainty propagation of UAV motion into RIS channel quality. Because an EKF is used to estimate the UAV's orientation, variances of the measured orientation variables are available. This orientation data enables us to assess the reliability of the estimated RIS channels through uncertainty propagation, providing a basis for informed, efficient data processing and optimal RIS configuration selection. We derived key quantities to characterize UAV orientation uncertainty, applying the Guide to the Expression of Uncertainty in Measurement (GUM) framework to quantify its effects the overall channel link quality. We demonstrate that the uncertainty area around the channel link can be significantly reduced, when transforming the uncertainly covariance matrix into an annular section. Simulations show how uncertainty propagation enables EKF-based RIS channel estimations to stay within confidence intervals, even with changing RIS configurations. This validates UAV-mounted RIS as a real-time capable solution for robust and adaptive communications despite environmental disturbances.


\bibliographystyle{IEEEtran}
\small
\bibliography{bibliography}
\balance
\end{document}